\documentclass[cup5b]{caps}
\usepackage{graphicx}
\usepackage{amssymb}
\usepackage{ociwsymp3e}  

\HeadText{C. R. Mullis et al.}  

\def\plotone#1{\centering \leavevmode
\includegraphics[width=.95\columnwidth]{#1}}

\def\plotone#1{\centering \leavevmode
\includegraphics[width=.95\columnwidth]{#1}}

\begin{document}

\pagenumbering{arabic}

\author[]{C. R. MULLIS$^{1}$, B. R. MCNAMARA$^{2}$, H. QUINTANA$^{3}$, A. VIKHLININ$^{4}$,\\ J. P. HENRY$^{5}$, I. M. GIOIA$^{6}$, A. HORNSTRUP$^{7}$, W. FORMAN$^{4}$, and C. JONES$^{4}$ \\
(1) European Southern Observatory, Germany, (2) Dept. of Physics \& Astronomy, Ohio Univ., USA,\\ (3) Dept. de Astronomia, Pontificia Univ. Catolica de Chile, (4) Harvard-Smithsonian Center for\\ Astronomy, USA, (5) Inst. for Astronomy, Univ. of Hawai`i, USA, (6) Istituto di Radioastronomia del \\CNR, Italy, (7) Danish Space Research Inst., Copenhagen, Denmark}

\chapter{The 160 deg$^{2}$ {\em ROSAT\/} Survey\\
Revised Catalog \& Cluster Evolution
}


\section{Overview of the Survey}

We have constructed a large, statistically complete sample of galaxy
clusters serendipitously detected as extended X-ray sources in 647
{\em ROSAT\/} PSPC pointed observations (Vikhlinin et al.\ 1998a,
hereafter V98). The survey covers 158 square degrees with a median
sample flux limit of 1.2 $\times$ 10$^{-13}$ \mbox{erg cm$^{-2}$ s$^{-1}$}
(0.5--2.0 keV). Our sample consists of 201 clusters of galaxies
characterized by a median redshift of z=0.25 and a maximum of
z=1.26. With 22 clusters at z > 0.5, the 160 Square Degree {\em
ROSAT\/} Survey (hereafter 160SD) is the largest high-redshift sample
of X-ray-selected clusters published to date.

\begin{figure}
\includegraphics[width=4in,angle=0]{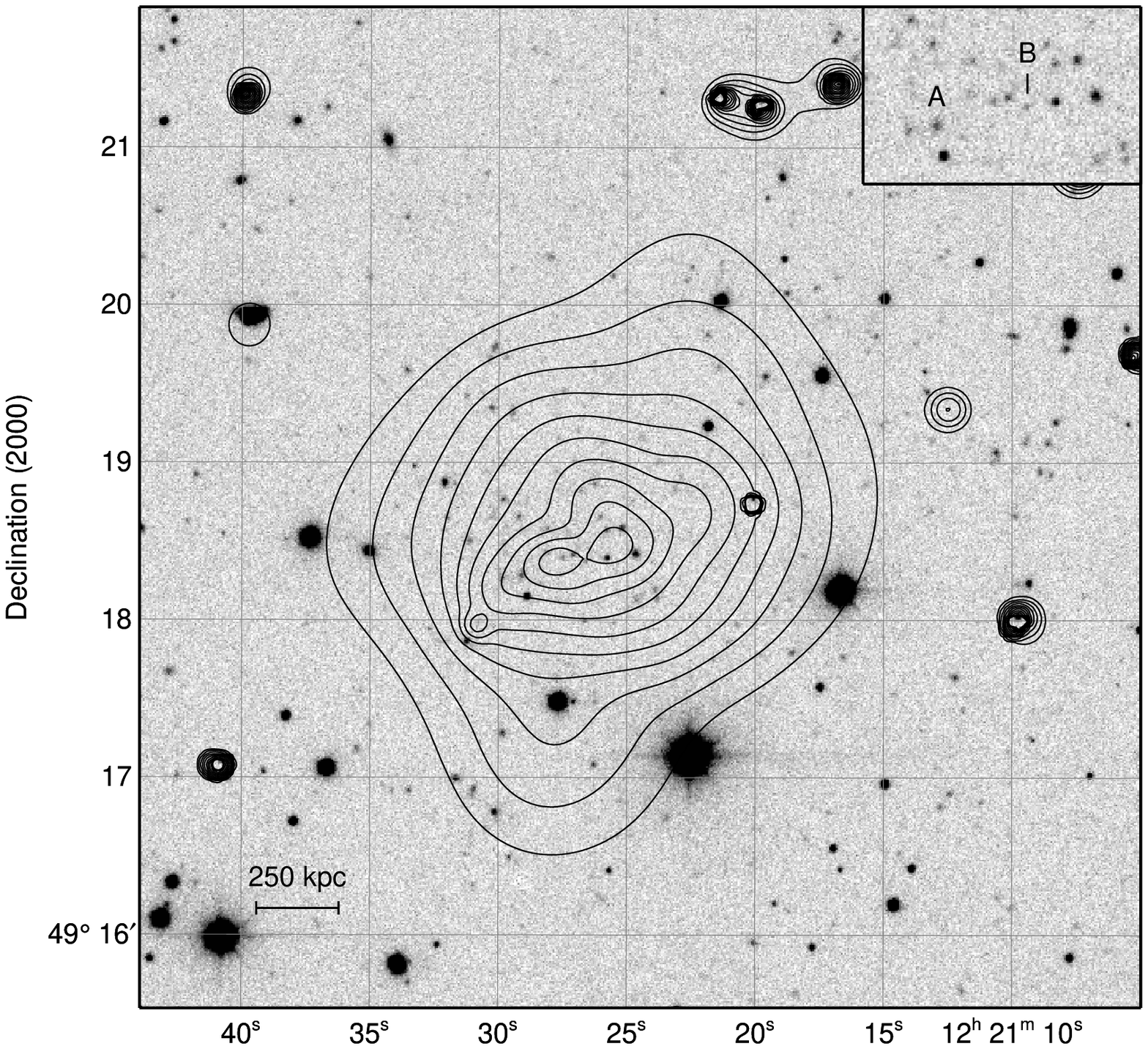}
\includegraphics[width=4in,height=3in,angle=0]{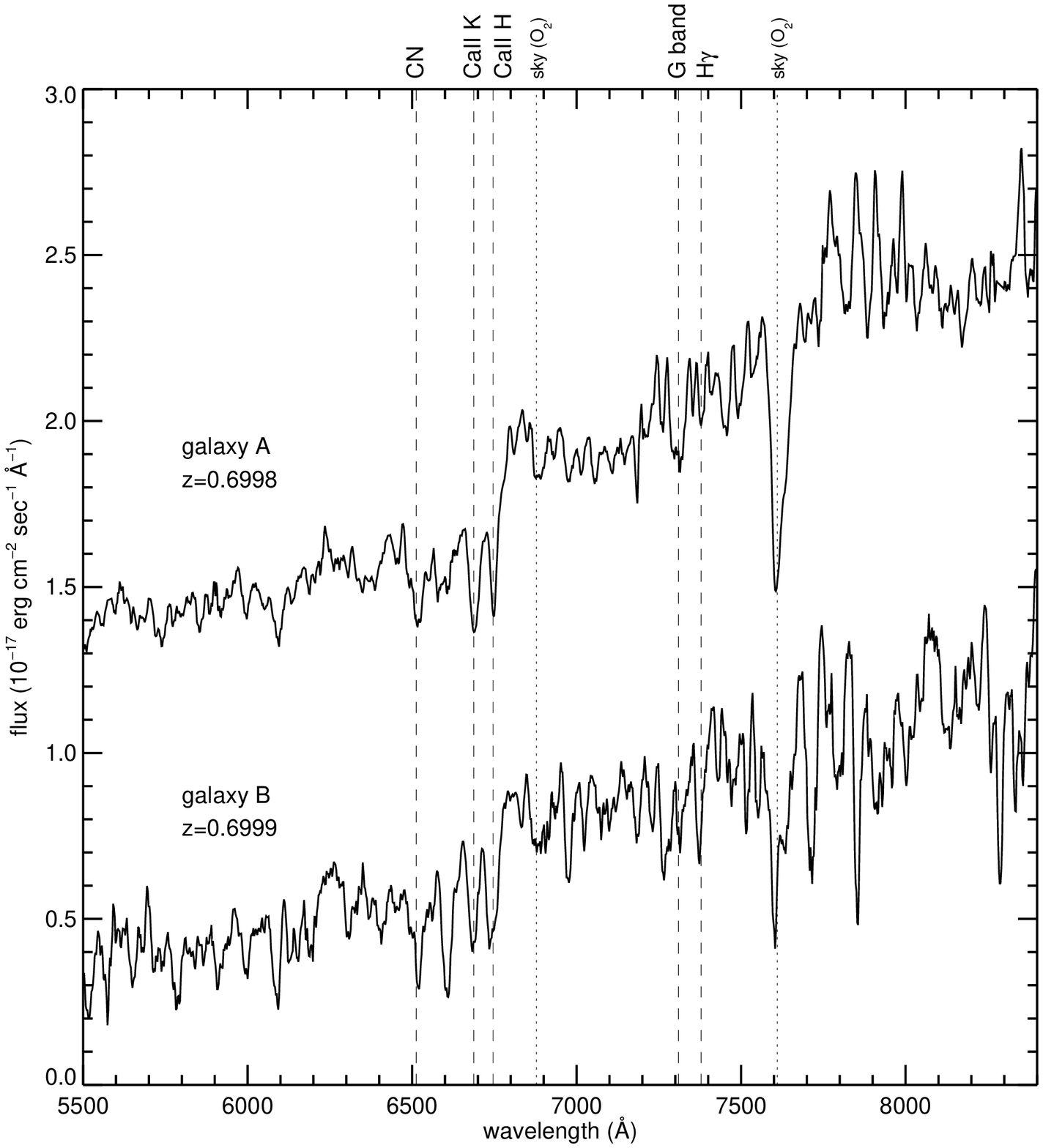}
\caption{RX\,J1221.4+4918 (\#119): a distant cluster at $z=0.700$. A 5
minute $R$-band image (upper panel) taken with the FLWO 1.2m is
overlaid with adaptively smoothed X-ray flux contours in the
0.7--2.0\,keV band from an 80\,ks observation with the {\em Chandra}
ACIS-I.  Contours are logarithmically spaced by factors of 1.4 with
the lowest contour a factor of 2 above the background ($5.5 \times
10^{-4}$ counts s$^{-1}$ arcmin$^{2}$).  The inset
 indicates the cluster galaxies for which redshifts were
measured using longslit spectra from Keck-II
LRIS (lower panel). 
\label{fig:RXJ1221.4+4918}}
\end{figure}

The galaxy clusters of our survey were initially confirmed in a
comprehensive program of optical imaging of all the candidates (e.g.,
CCD image in \mbox{Figure \ref{fig:RXJ1221.4+4918}}) and optical
spectroscopy for a limited subsample. Two-thirds of the cluster
redshifts in the original catalog (V98) were photometric estimates and
the remaining third were spectroscopic measurements. We recently
completed spectroscopy of the entire sample, and we now have
spectroscopic redshifts for 99.5\% of the clusters (e.g., spectra in
\mbox{Figure \ref{fig:RXJ1221.4+4918}}; Mullis et al.\ 2003a). Details of
the revised cluster sample are discussed in the next section.

The 160SD clusters have been used to study the evolution of cluster
X-ray luminosities and radii (Vikhlinin et al.\ 1998b), to present
evidence for a new class of X-ray overluminous elliptical galaxies or
"fossil groups" (Vikhlinin et al.\ 1999), to analyze the correlation
of optical cluster richness with redshift and X-ray luminosity
(McNamara et al.\ 2001), and to discover a wide-angle gravitational
lens (Munoz et al.\ 2001). Chandra observations of high-redshift 160SD
clusters have been used to make an accurate determination of the
evolution of the scaling relations between X-ray luminosity,
temperature, and gas mass (Vikhlinin et al.\ 2002), and to derive
cosmological constraints from the evolution of the cluster baryon mass
function (Vikhlinin et al.\ 2003b). Subsequently we will describe the
most recent results of our survey which concern the evolution of the
number density of clusters.

We assume an Einstein-\mbox{de Sitter} cosmological model
with $H_{0} = 50$ h$_{\rm 50}$ km s$^{-1}$ Mpc$^{-1}$ and $\Omega_{M}=1$
($\Omega_{\Lambda} = 0$), and quote \mbox{X-ray} fluxes and
luminosities in the {\mbox 0.5--2.0 keV} energy band.

\section{Revised Cluster Sample with Spectroscopic Redshifts}

Since the initial follow-up observations we have gone on to measure
spectroscopic redshifts for 110 additional clusters from our 160SD
survey using the Keck-II 10m and the University of Hawai`i (UH) 2.2m
telescopes at the Mauna Kea Observatories, and the ESO 3.6m telescope
at La Silla Observatory. Combining these new
redshifts with 76 measurements reported by V98 and 14 redshifts from
the literature and private communications results in essentially
complete (200 of 201 clusters) spectroscopic coverage for our entire
sample. In general the photometric redshifts
were quite reliable. The membership of the 160SD cluster sample has
proven to be remarkably stable since first presented by V98. The
revised sample consists of 201 confirmed clusters, 21 false
detections, and one source obscured by Arcturus.

\section{Cluster Evolution}

\begin{figure}
\plotone{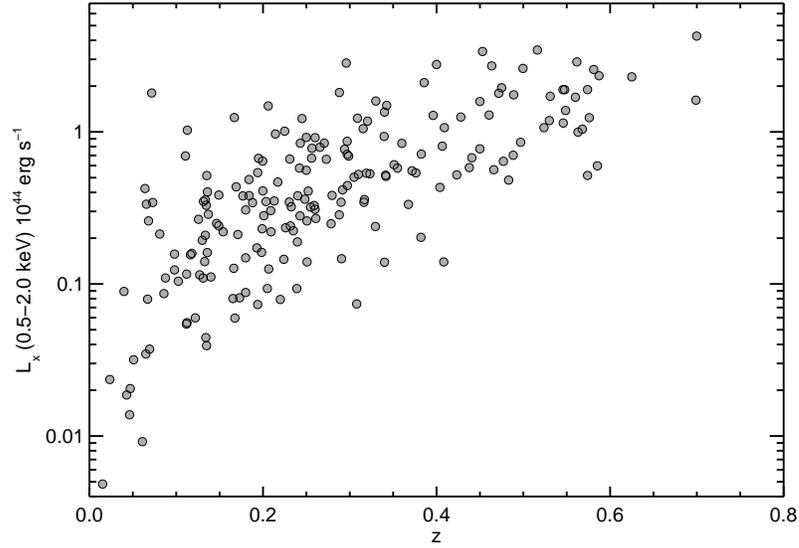}
\caption{X-ray luminosity and redshift distribution of the 160SD
cluster sample.  The median redshift is $z_{\rm median}=0.25$ and the
median X-ray luminosity is $L_{\rm X,median} = 4.2 \times 10^{43}$
\mbox{erg s$^{-1}$}.  
\label{fig:lxz}}
\end{figure}

\begin{figure}
\plotone{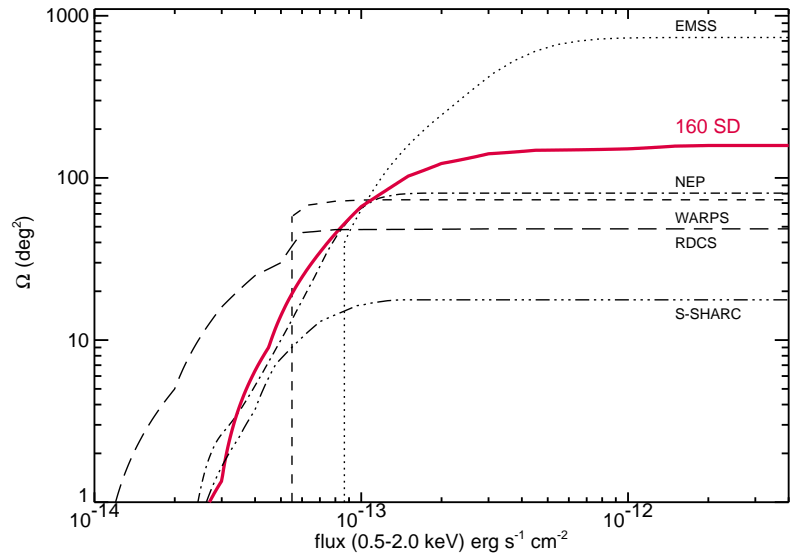}
\caption{The sky coverage of various {\em ROSAT\/} cluster surveys along
with the EMSS. The survey selection of the 160SD survey is
highlighted. References for the surveys: EMSS, Henry et al.\
1992; NEP, Mullis 2001; WARPS, H. Ebeling 2000, private comm.;
RDCS, Rosati et al.\ 1995; S-SHARC, Burke et al.\ 1997.
\label{fig:skycov}}
\end{figure}

We have the three requirements for testing evolution in the number
density of clusters. First we have a statistically complete sample
which probes sufficiently high luminosity and redshift (\mbox{Figure
\ref{fig:lxz}}) where differential evolution should be strongest. Second
we have accurately measured the survey selection function via
extensive Monte Carlo simulations (\mbox{Figure \ref{fig:skycov}}) which
permits us to compute reliable search volumes necessary for
constructing volume-normalized diagnostics. And finally, the local
X-ray luminosity function (XLF), which forms the baseline for the
no-evolution scenario, is well determined.

\begin{figure}
\includegraphics[width=4in,height=3in]{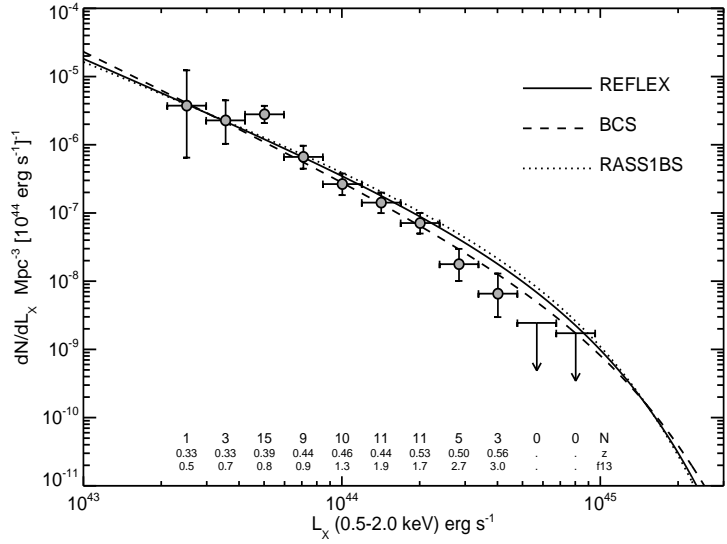}
\caption{High-redshift 160SD cluster XLF ($z=0.3$--0.8). Note the
deficit of clusters above ~3 $\times$ 10$^{44}$ \mbox{erg s$^{-1}$}. In this luminosity-redshift
interval 3 clusters are detected where 15 (20, 24) clusters are
predicted using the BCS (REFLEX, RASS1BS) XLF as the no-evolution
baseline.
\label{fig:xlf_hiz}}
\end{figure}

\begin{figure}
\includegraphics[width=4in,height=3in]{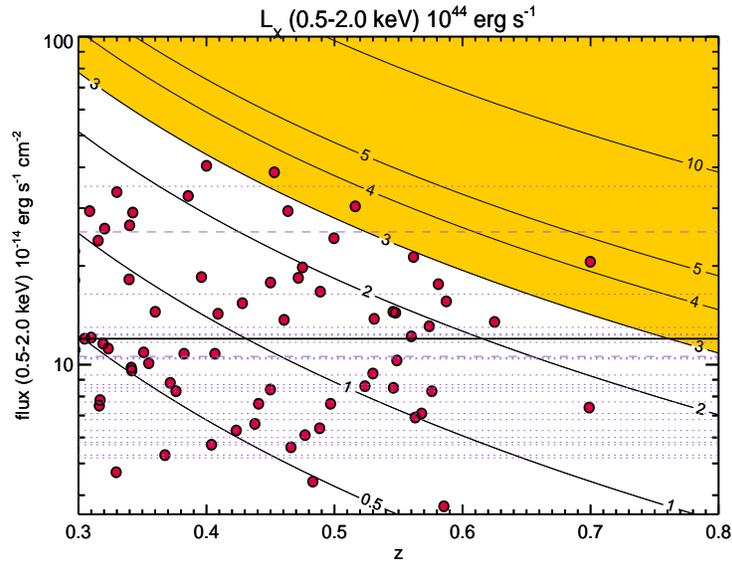}
\caption{X-ray flux versus redshift for the 73 clusters of the 160SD
sample at \mbox{$z > 0.3$}. Isocontours of X-ray luminosity (10$^{44}$ \mbox{erg s$^{-1}$}) are
overplotted. The luminosity regime of the observed cluster deficit is
 shaded. The "missing" clusters would be high-flux
sources. The flux of the 21
false detections of the 160SD survey are represented by the dotted
horizontal lines.
\label{fig:false}}
\end{figure}

We present the high-redshift XLF in \mbox{Figure \ref{fig:xlf_hiz}}. At
lower luminosities the abundance is in good agreement with the local
value. However, at higher luminosities (>3 $\times$ 10$^{44}$ \mbox{erg s$^{-1}$}) there is a
systematic departure from the no-evolution curve.  We can quantify
this cluster deficit by comparing the number of detected clusters
versus the expected number computed by folding the local XLF through
our selection function. Between $0.3 < z < 0.8$ and above \mbox{3 $\times$ 10$^{44}$,} we
find 3 clusters where 15 (20, 25) clusters are predicted using the BCS
(REFLEX, RASS1BS) XLF as the no-evolution baseline. This deficit is
significant at the 3.5 (4.5, 5.2) sigma level. In \mbox{Figure
\ref{fig:false}} we show that the false detections can not resolve this
mismatch because the implied luminosities are too low.

\end{document}